\pdfoutput=1

\RequirePackage[l2tabu, orthodox]{nag}

\documentclass[10pt, a4paper, conference]{IEEEtran}

\usepackage[pass]{geometry}
\usepackage[utf8]{inputenc}
\usepackage{amsmath, amsfonts}
\usepackage{newtxtext, newtxmath}
\usepackage[T1]{fontenc}
\usepackage{mathtools}
\usepackage{microtype}
\usepackage{booktabs}
\usepackage[style=ieee]{biblatex}
\usepackage{graphicx}
\usepackage{subcaption}
\usepackage{tikz, pgfplots}
\usepackage[per-mode=symbol, binary-units]{siunitx}
\usepackage{xfrac}
\usepackage{xparse}
\usepackage{xspace}
\usepackage{xcolor}
\usepackage{algorithm}
\usepackage{algpseudocode} 
\usepackage{orcidlink}
\usepackage{url}
\usepackage{hyperref}

\newcommand\ie{i.\,e.\@\xspace}

\newcommand\eg{e.\,g.\@\xspace}

\newcommand\etal{et al.\@\xspace}

\newcommand\discoroute{\textsc{DisCoRoute}\xspace}

\definecolor{tab-indigo}{HTML}{334499}
\definecolor{tab-cyan}{HTML}{88CCEE}
\definecolor{tab-teal}{HTML}{44AA99}
\definecolor{tab-green}{HTML}{117733}
\definecolor{tab-sand}{HTML}{CCBB77}
\definecolor{tab-rose}{HTML}{DD6677}
\definecolor{tab-wine}{HTML}{882255}
\definecolor{tab-purple}{HTML}{AA4499}
\definecolor{tab-gray}{HTML}{AAAAAA}

\pgfplotsset{compat=1.17}
\usetikzlibrary{angles, arrows.meta, calc, decorations.text, decorations.pathreplacing, decorations.pathmorphing, intersections, math, quotes}

\DeclareMathOperator{\sgn}{sgn}

\algrenewcommand\alglinenumber[1]{\scriptsize #1}
\algrenewcommand\algorithmiccomment[1]{\hfill{\color{tab-gray}$\triangleright$ #1}}%

\newtheorem{thm}{Theorem}
\newtheorem{conj}{Conjecture}

\tikzmath{
    function is_ascending_deg(\u) { return (-90 <= \u) && (\u <= 90); };
    function is_ascending(\u) { return (-pi/2 <= \u) && (\u <= pi/2); };
    function get_lat_deg(\a,\u,\Lzero) { return asin(sin(\a) * sin(\u)); };
    function get_lat(\a,\u,\Lzero) { return rad(get_lat_deg(deg(\a), deg(\u), deg(\L0))); };
    function get_lon_deg(\a,\u,\Lzero) {
        \res = \Lzero + atan(cos(\a) * tan(\u));
        if is_ascending_deg(\u) == 0 then { \res = \res + 180; };
        if \res > 180 then { \res = \res - 360; } else {
            if \res < -180 then { \res = \res + 360; };
        };
        return \res;
    };
    function get_lon(\a,\u,\Lzero) { return rad(get_lon_deg(deg(\a), deg(\u), deg(\L0))); };
    function euclidean3d_deg(\aA,\uA,\LzeroA,\aB,\uB,\LzeroB) {
        \latA = get_lat_deg(\aA,\uA,\LzeroA);
        \latB = get_lat_deg(\aB,\uB,\LzeroB);
        \lonA = get_lon_deg(\aA,\uA,\LzeroA);
        \lonB = get_lon_deg(\aB,\uB,\LzeroB);
        \XA = cos(\latA) * cos(\lonA);
        \XB = cos(\latB) * cos(\lonB);
        \YA = cos(\latA) * sin(\lonA);
        \YB = cos(\latB) * sin(\lonB);
        \ZA = sin(\latA);
        \ZB = sin(\latB);
        return sqrt((\XA-\XB)^2 + (\YA-\YB)^2 + (\ZA-\ZB)^2);
    };
    function isl_distance_deg(\a,\u,\Lzero,\deltaF,\deltaO) {
        \LzeroB = \Lzero + \deltaO;
        \uB = \u + \deltaF;
        if \uB > 180 then { \uB = \uB - 360; } else {
            if \uB < -180 then { \uB = \uB + 360; };
        };
        return euclidean3d_deg(\a,\u,\Lzero,\a,\uB,\LzeroB);
    };
}

\title{Distributed On-Demand Routing for LEO Mega-Constellations: A Starlink Case Study}

\author{%
    \IEEEauthorblockN{Gregory Stock \orcidlink{0000-0001-5170-2019} \qquad Juan A. Fraire \orcidlink{0000-0001-9816-6989} \qquad Holger Hermanns \orcidlink{0000-0002-2766-9615}}
    \IEEEauthorblockA{%
        Saarland University -- Computer Science, 
        Saarland Informatics Campus, 66123 Saarbrücken, Germany\\
        \{\href{mailto:stock@depend.uni-saarland.de}{stock}, \href{mailto:juanfraire@depend.uni-saarland.de}{juanfraire}, \href{mailto:hermanns@depend.uni-saarland.de}{hermanns}\}@depend.uni-saarland.de
    }
}

\hypersetup{
    pdftitle={Distributed On-Demand Routing for LEO Mega-Constellations: A Starlink Case Study},
    pdfauthor={Gregory Stock; Juan A. Fraire; Holger Hermanns}
}

\begin{document}

\maketitle

\thispagestyle{plain}
\pagestyle{plain}

\begin{abstract}
    The design and launch of large-scale satellite networks create an imminent demand for efficient and delay-minimising routing methods.
    With the rising number of satellites in such constellations, pre-computing all shortest routes between all satellites and for all times becomes more and more infeasible due to space and time limitations.
    Even though distributed on-demand routing methods were developed for specific LEO satellite network configurations, they are not suited for increasingly popular mega-constellations based on Walker Delta formations.

    The contributions of this paper are twofold.
    First, we introduce a formal model that mathematically captures the time-evolving locations of satellites in a Walker Delta constellation and use it to establish a formula to compute the minimum number of ISL hops between two given satellites.
    In the second part, we present an on-demand hop-count-based routing algorithm that approximates the optimal path while achieving superior performance compared to classical shortest-path algorithms like Dijkstra.
\end{abstract}

\medskip
\noindent
\emph{This is an extended version of a paper published in ASMS/SPSC 2022~\cite{Stock22ASMS} containing proofs and further details.}

\section{Introduction}

Recent technology advances have led to a rapidly increasing number of satellite launches into low Earth orbit (LEO).
Most of these satellites are part of huge mega-constellations that aim at providing global and fast communication, \eg for internet access.
Since most constellations scheduled for future deployment will rely on inter-satellite links (ISLs), packets will usually need to take a high number of hops when travelling from source to destination.
The huge number of satellites adds to the complexity of routing in such a network.
There is thus a need for specialised and efficient routing algorithms, especially since space networks have to compete with existing and already well-understood ground networks where the topology is more stable compared to their counterparts in space.
Besides offering global coverage, satellite operators aim at providing higher speeds and lower latencies compared to traditional broadband.
For this, it is critical that one can quickly determine routes that minimise the delay.
Part of the solution lies already in the design phase of the constellation where a lot of room for optimisation exists.
Routing, however, remains an important problem once the constellation is in orbit and operable.
Packets need to travel large distances in space and pass several satellites that relay the traffic from one satellite to its neighbour.
It is, therefore, generally beneficial to select short routes.
However, since every extra ISL relay adds additional processing costs, it is even more important to keep the number of hops as low as possible.
This raises the important challenges of how to calculate the number of hops needed to connect two satellites in the constellation and how to use these results when it comes to routing.

This paper first provides a mathematical model to express and work with satellite positions in a Walker Delta constellation, a popular constellation type consisting of two locally separate overlapping meshes, an ascending and a descending one (\autoref{sec:model}).
Then, this model is used in \autoref{sec:min-hop-count} to derive a formula for the minimum number of ISL hops required to connect two satellites.
These results are the basis for two new on-demand routing algorithms that produce routes with the minimum possible number of hops (\autoref{sec:routing}).
The algorithms have almost no overhead and work in a distributed way, as neither an exhaustive exploration nor a centralised pre-computation of the path is needed.
The first algorithm simply selects hops probabilistically while assuring hop-minimality.
The second algorithm uses a heuristic to generate near-optimal approximations of the best route but still comes at almost no cost.
This heuristic is based on observations on the distances spanned by inter- and intra-plane hops that we introduce in \autoref{sec:hop-distances}.
It is optimal, \ie it is guaranteed to find the overall shortest route, for the base case of constellation parameters.
Finally, we report (\autoref{sec:evaluation}) on a thorough empirical evaluation spanning simulations of artificial as well as real-world constellations, with a focus on Starlink.
The proposed algorithms are significantly faster than the state of the art while maintaining competitiveness regarding the quality of the routes.
We come to the conclusion that the hop-count-based algorithms are superior to classical routing algorithms based solely on shortest paths.

\section{Background}\label{sec:model}

\subsection{Existing Routing Algorithms}

Routing, in general, is a well-studied topic that is also addressed by many research papers in the context of space surveyed in~\cite{DBLP:journals/jcin/QiMW0H16, Madni20}.
Yet, most algorithms have been developed quite some time ago and have had the needs of those times as motivations.
For example, they only support polar constellations (\eg Walker Star) or are designed for small ATM-like packets which have minimal relay time so that the hop count is not so important~\cite{Li06}.
This means that these algorithms are not necessarily suitable or optimal for upcoming mega-constellations.
This paper focuses on a type of constellation called \emph{Walker Delta} that has recently found popularity in the design of upcoming mega-constellations according to regulatory filings.
The satellites in such a constellation follow circular orbits and are partitioned into different orbital planes.
Before we mathematically describe these constellations in more detail, we first consider the modelling of individual satellites.

\subsection{Satellite Model}

\begin{figure}
    \centering%
    \begin{tikzpicture}[]
    \pgfmathsetmacro{\r}{2}
    \coordinate (O) at (0, 0);

    \filldraw[draw=tab-wine, fill=tab-wine!20] (0, -\r) arc [start angle=-90, end angle=90, radius=\r];
    \filldraw[draw=tab-teal, fill=tab-teal!20] (0, \r) arc [start angle=90, end angle=270, radius=\r];

    \coordinate (u0) at (\r, 0);
    \coordinate (u90) at (0, \r);
    \coordinate (u180) at (-\r, 0);
    \coordinate (u270) at (0, -\r);

    \foreach \x in {0,...,7} {
        \coordinate (S\x) at ({\r * cos(\x * 45)}, {\r * sin(\x * 45)});
    }

    \coordinate (Sat) at ({\r * cos(66)}, {\r * sin(66)});

    \draw[thick, draw=tab-wine] (S0) -- ++(0.1, 0) node[right] {$0$};
    \draw[thick, draw=tab-wine] (S1) -- ++(0.07, 0.07) node[above right] {$\frac{\pi}{4}$};
    \draw[thick, draw=tab-wine] (S2) -- ++(0, 0.1) node[above] {$\frac{\pi}{2}$};
    \draw[thick, draw=tab-teal] (S3) -- ++(-0.07, 0.07) node[above left] {$\frac{3\pi}{4}$};
    \draw[thick, draw=tab-teal] (S4) -- ++(-0.1, 0) node[left] {$\pm\pi$};
    \draw[thick, draw=tab-teal] (S5) -- ++(-0.07, -0.07) node[below left] {$-\frac{3\pi}{4}$};
    \draw[thick, draw=tab-wine] (S6) -- ++(0, -0.1) node[below] {$-\frac{\pi}{2}$};
    \draw[thick, draw=tab-wine] (S7) -- ++(0.07, -0.07) node[below right] {$-\frac{\pi}{4}$};

    \fill (Sat) circle (2pt);
    \draw[dotted, tab-wine] (O) -- (S0);
    \draw[dotted, tab-wine] (O) -- (S2);
    \draw[dotted, tab-teal] (O) -- (S4);
    \draw[dotted, tab-wine] (O) -- (S6);
    \draw[dashed] (O) -- (Sat);
    \pic["$u$", draw, angle radius=7mm, ->, >=stealth] {angle = S0--O--Sat};

    \path[decorate, decoration={text along path, text={ascending segment}, text align=center}] (0, -1.8) arc [start angle=-90, end angle=90, radius=1.8];
    \path[decorate, decoration={text along path, text={descending segment}, text align=center}] (0, 1.8) arc [start angle=90, end angle=270, radius=1.8];

    \fill (O) circle (3pt);
\end{tikzpicture}%
    \caption{Frontal view on a circular orbit, showing the argument of latitude~$u$ and the two regions of flying direction.}
    \label{fig:phase-angle}
\end{figure}
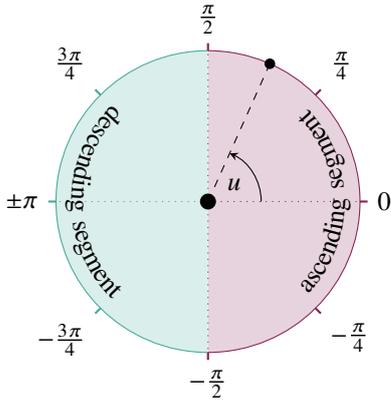

The current position of a satellite is typically represented using the six Keplerian elements, also known as classical orbital elements.
Most important are the \emph{longitude of the ascending node}~$\Omega$ and the \emph{true anomaly}~$\nu$.
In the following, we will usually refer to the \emph{initial longitude of the ascending node}~$L_0\in \mathopen[-\pi, \pi\mathclose[$ at some Epoch, which is a constant and independent of the current time~$t$.
The current longitude of the ascending node~$\Omega$ at time $t$ is given by $\Omega = L_0 - \omega_E\cdot t$, where $\omega_E$ is the angular speed of Earth's rotation.
Since the orbits are circular (\ie eccentricity $e=0$), the true anomaly is undefined, as the periapsis cannot be uniquely determined.
Therefore, the \emph{argument of latitude} (or phase angle)~$u \in \mathopen[-\pi, \pi\mathclose[$ is used instead, which is the angle between the ascending node and the satellite and basically defines the position of the satellite in its orbit.
We say that a satellite is \emph{ascending}, \ie flying in north-east direction, if $u\in\mathopen[-\frac{\pi}{2}, \frac{\pi}{2}\mathclose]$ and call it otherwise \emph{descending}, \ie flying towards the south-east (see \autoref{fig:phase-angle}).
The remaining orbital elements, \ie \emph{semi-major axis}~$a$ and \emph{inclination}~$\alpha$, are equal for all satellites and therefore considered global constants of the constellation.
Note that the semi-major axis of the orbit is equal to its radius due to its circular shape.
Further, note that we define a satellite's \emph{altitude}~$h$ relative to the surface of the Earth, \ie the orbit radius is equal to the sum of $h$ and Earth's semi-major axis $r_a = \SI{6378.137}{\km}$ as specified in the WGS84 reference system~\cite{WGS84}.

While it is very descriptive and convenient to model the position of a single satellite using these basic orbital parameters, it is usually not very useful when comparing the positions of different satellites, \eg calculating their distance.
Therefore, we show in the following how the argument of latitude~$u$ and the (initial) longitude of the ascending node~$L_0$ can be converted first to geodetic coordinates (latitude \& longitude) and then to a cartesian coordinate system known as ECEF (Earth-centered, Earth-fixed).

\subsubsection{Keplerian to Geodetic}

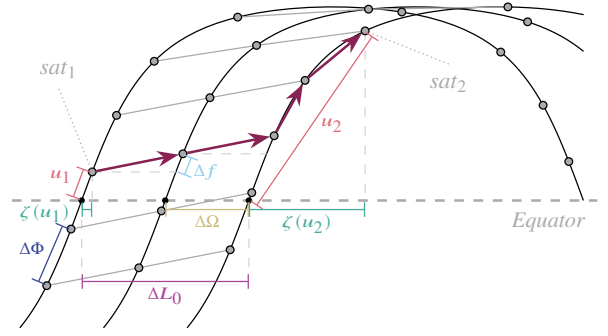
\begin{figure}
    \centering%
    \pgfdeclaredecoration{simple line}{initial}{
  \state{initial}[width=\pgfdecoratedpathlength-1sp]{\pgfmoveto{\pgfpointorigin}}
  \state{final}{\pgflineto{\pgfpointorigin}}
}
\tikzset{
   shift left/.style={decorate, decoration={simple line, raise=#1}},
   shift right/.style={decorate, decoration={simple line, raise=-1*#1}},
}
\begin{tikzpicture}[scale=1.2]
    \tikzmath{
        \inc = 70;
        \domega = 30;
        \dphi = 22;
        \df = 7;
        \uorigin = -33;
    }
    \begin{axis}[width=9cm, unit vector ratio={1, 1}, samples=100, domain=-60:125, grid=none, hide x axis=true, hide y axis=true]
        \draw[tab-gray, thick, dashed] (-55, 0) -- (155, 0) node [below left, font=\scriptsize] {\emph{Equator}};

        \addplot[black, domain=-50:180] plot ({get_lon_deg(\inc, \x, -\domega)}, {get_lat_deg(\inc, \x, -\domega)});
        \addplot[black, domain=-50:120.7] plot ({get_lon_deg(\inc, \x, 0)}, {get_lat_deg(\inc, \x, 0)});
        \addplot[black, domain=-50:101.2] plot ({get_lon_deg(\inc, \x, \domega)}, {get_lat_deg(\inc, \x, \domega)});

        \pgfplotsinvokeforeach{0,...,9} {
            \coordinate (A#1) at ({get_lon_deg(\inc, \uorigin+#1*\dphi, -\domega)}, {get_lat_deg(\inc, \uorigin+#1*\dphi, -\domega)});
        }
        \pgfplotsinvokeforeach{0,...,6} {
            \coordinate (B#1) at ({get_lon_deg(\inc, \uorigin+\df+#1*\dphi, 0)}, {get_lat_deg(\inc, \uorigin+\df+#1*\dphi, 0)});
        }
        \pgfplotsinvokeforeach{0,...,5} {
            \coordinate (C#1) at ({get_lon_deg(\inc, \uorigin+2*\df+#1*\dphi, \domega)}, {get_lat_deg(\inc, \uorigin+2*\df+#1*\dphi, \domega)});
        }

        \coordinate (EA) at (-\domega, 0);
        \coordinate (EB) at (0, 0);
        \coordinate (EC) at (\domega, 0);

        \coordinate (ECA2) at ($(EC -| A2) + (0, -20) - (0, 0)$);
        \coordinate (ECC4) at ($(EC -| C4) + (0, -20) - (0, 0)$);
        \draw[tab-gray!50, dashed, thin] (A2) -- ({get_lon_deg(\inc, \uorigin+2*\dphi, 0)}, {get_lat_deg(\inc, \uorigin+2*\dphi, 0)});
        \draw[tab-gray!50, dashed, thin] (B2) -- ({get_lon_deg(\inc, \uorigin+\df+2*\dphi, \domega)}, {get_lat_deg(\inc, \uorigin+\df+2*\dphi, \domega)});
        \draw[tab-gray!50, dashed, thin] (A2) -- (EC -| A2);
        \draw[tab-gray!50, dashed, thin] (C4) -- (EC -| C4);
        \draw[tab-gray!50, dashed, thin] (EA) -- ($(EA)+(0, -26)-(0, 0)$);
        \draw[tab-gray!50, dashed, thin] (EC) -- ($(EC)+(0, -26)-(0, 0)$);

        \pgfplotsinvokeforeach{0,...,5} {
            \draw[tab-gray, thin] (A#1) -- (B#1);
            \draw[tab-gray, thin] (B#1) -- (C#1);
        }

        \pgfplotsinvokeforeach{0,...,9} {\draw [black, fill=tab-gray] (A#1) circle (1.2pt);}
        \pgfplotsinvokeforeach{0,...,6} {\draw [black, fill=tab-gray] (B#1) circle (1.2pt);}
        \pgfplotsinvokeforeach{0,...,5} {\draw [black, fill=tab-gray] (C#1) circle (1.2pt);}

        \pgfplotsinvokeforeach{A,B,C} {\fill (E#1) circle (1pt);}

        \draw[tab-rose, |-|, shift left=1mm] (EA) -- node[left, xshift=-1pt, yshift=3pt, font=\tiny] {$u_1$} (A2);
        \draw[tab-rose, |-|, shift right=1mm] (EC) -- node[right, xshift=1pt, yshift=-3pt, font=\tiny] {$u_2$} (C4);
        \draw[tab-indigo, |-|, shift left=1mm] (A0) -- node[left, xshift=-1pt, yshift=3pt, font=\tiny] {$\Delta\Phi$} (A1);
        \draw[tab-sand, |-|, shift right=1mm] (EB) -- node[below, yshift=-1pt, font=\tiny] {$\Delta\Omega$} (EC);
        \draw[tab-cyan, |-|, shift right=1mm] ({get_lon_deg(\inc, \uorigin+2*\dphi, 0)}, {get_lat_deg(\inc, \uorigin+2*\dphi, 0)}) -- node[right, xshift=1pt, yshift=-3pt, font=\tiny] {$\Delta f$} (B2);
        \draw[tab-teal, |-|, shift right=1mm] (EC) -- node[below, yshift=-1pt, font=\tiny] {$\zeta(u_2)$} (EC -| C4);
        \draw[tab-teal, |-|, shift right=1mm] (EA) node[left, yshift=-4pt, font=\tiny] {$\zeta(u_1)$} -- (EA -| A2);
        \draw[tab-purple, |-|, shift right=1mm] ($(EA)+(0, -26)-(0, 0)$) -- node[below, yshift=-1pt, font=\tiny] {$\Delta L_0$} ($(EC)+(0, -26)-(0, 0)$);

        \draw[->, tab-wine, line width=0.8pt, >=Stealth, shorten <=1.4pt, shorten >=1.4pt] (A2) -- (B2);
        \draw[->, tab-wine, line width=0.8pt, >=Stealth, shorten <=1.4pt, shorten >=1.4pt] (B2) -- (C2);
        \draw[->, tab-wine, line width=0.8pt, >=Stealth, shorten <=1.4pt, shorten >=1.4pt] (C2) -- (C3);
        \draw[->, tab-wine, line width=0.8pt, >=Stealth, shorten <=1.4pt, shorten >=1.4pt] (C3) -- (C4);

        \draw[tab-gray, densely dotted, shorten >=1.2pt] ($(A2)+(-12, 30)-(0, 0)$) node[above, font=\scriptsize] {$\mathit{sat}_1$} -- (A2);
        \draw[tab-gray, densely dotted, shorten >=1.2pt] ($(C4)+(30, -12)-(0, 0)$) node[below, font=\scriptsize] {$\mathit{sat}_2$} -- (C4);
    \end{axis}
\end{tikzpicture}%
    \caption{Ground plot of a section of a constellation annotated by the various modelling parameters. All indicated parameters are given as angles measured from the centre of Earth.}
    \label{fig:basic-geometry}
\end{figure}

Geodetic coordinates are specified using \emph{latitude}~$\varphi$, \emph{longitude} $\lambda$, and the \emph{height}~$h$.
The conversion of a satellite position to geodetic coordinates is given by:
\begin{align*}
    \varphi & = \arcsin(\sin\alpha \cdot \sin u)   \begingroup\color{tab-gray}\in [-\alpha, \alpha]\endgroup                                                                          \\
    \lambda & = \mathcal{N}\big(L_0 - \omega_E\cdot t + \zeta(u)\big) = \mathcal{N}\big(\Omega + \zeta(u)\big) \begingroup\color{tab-gray}\in \mathopen[-\pi, \pi\mathclose[\endgroup
\end{align*}
Here, $\zeta(u)$ indicates the \emph{longitude difference} of a satellite to its ascending node (see \autoref{fig:basic-geometry}):
\begin{align*}
    \zeta(u) = \arctan(\cos\alpha \cdot \tan u) + \begin{cases}
                                                      0   & \textrm{asc. segment}  \\
                                                      \pi & \textrm{desc. segment}
                                                  \end{cases}
\end{align*}
$\mathcal{N}(x) = \big((x + \pi) \bmod 2\pi\big) - \pi$ is a normalisation function that ensures that the resulting values are within the desired interval $\mathopen[-\pi, \pi\mathclose[$.
An important application of geodetic coordinates is the calculation of the sub-satellite point.
It is defined as the point where a straight line from the centre of the Earth to the satellite intersects with the surface of the Earth.
Since the geodetic coordinate system is a spherical system, the sub-satellite point of a satellite has the same latitude and longitude as the satellite, \ie only its altitude differs.

\subsubsection{Geodetic to Cartesian}

The \emph{Earth-centered, Earth-fixed} coordinate system (ECEF) is a geocentric system that uses Cartesian coordinates.
This representation is well suited to compute distances between two objects in space.
Converting geodetic coordinates to Cartesian coordinates $(X, Y, Z)$ can be done as follows:
\begin{align*}
    \big(\underbrace{(r_a+h) \cos\varphi \cos\lambda}_{X},~\underbrace{(r_a+h) \cos\varphi \sin\lambda}_{Y},~\underbrace{(r_a+h) \sin\varphi}_{Z}\big)
\end{align*}

\subsection{Walker Delta Constellation}

A Walker Delta constellation consists of $P$ orbital planes that are evenly spaced around the Equator.
Each of these planes contains $Q$ evenly spaced satellites.
All satellites follow a circular orbit with the same inclination~$\alpha$ and altitude~$h$.
A Walker Delta constellation is often formally described by $\alpha:PQ/P/F$ where $F$ indicates the relative spacing between satellites in adjacent planes (see below).
Throughout this paper, we use $(o, i)$ to denote the $i$-th satellite in orbital plane~$o$.

The \emph{RAAN difference} (\ie difference in right ascension of the ascending node) between adjacent planes~$\Delta\Omega = \frac{2\pi}{P} \in [0, 2\pi]$ specifies how far apart neighbouring planes are from each other and only depends on the number of orbital planes.
Within an orbital plane, the \emph{phase difference}, \ie the difference in argument of latitude, between adjacent satellites $\Delta\Phi = \frac{2\pi}{Q} \in [0, 2\pi]$ can be computed from the number of satellites per plane.
Finally, the \emph{phase offset}~$\Delta f = \frac{2\pi F}{P Q} \in \mathopen[0, 2\pi\mathclose[$ between satellites in adjacent planes specifies the difference in argument of latitude between two horizontal neighbours.
This value is usually given in terms of a \emph{phasing factor}~$F \in \{0,\ldots,P-1\}$.
In contrast to arbitrary phase offsets, this factor ensures that the sum of phase offsets for all $P$ planes, \ie $P\cdot \Delta f$, is always a multiple of $\Delta\Phi$.
This property is required for the constellation to be symmetric, \ie to have the same $\Delta f$ for every satellite.
See \autoref{fig:basic-geometry} for a visualisation of these parameters.
Satellite $(o, i)$ can now be described by $\big(L_0 = \mathcal{N}(o\cdot\Delta\Omega),~u =\mathcal{N}(o\cdot\Delta f + i\cdot \Delta\Phi)\big)$ for $ 0 \leq o < P$ and $0 \leq i < Q$.

\paragraph{Inter-Satellite Links}

Throughout this paper, we assume that each satellite can establish four ISL links to its immediate neighbours, \ie an intra-plane link to the successor and predecessor in its own orbital plane and an inter-plane link each to the neighbour in the left and right plane.
Walker Delta constellations are typically deployed in low-inclination orbits (\eg $\ang{53.2}$ for Starlink's inner shell) where the Doppler impairments in cross-plane links can be coped with at the highest latitudes.
Formally, the predecessor and successor of satellite $(o, i)$ are given by $(o,~(i - 1) \bmod Q)$ and $(o,~(i + 1) \bmod Q)$, respectively.
The left neighbour is $(o - 1,~i)$ if $o \neq 0$ and $(P - 1,~(i - F) \bmod Q)$ otherwise.
Analogously, the right neighbour is $(o + 1,~i)$ if $o \neq P - 1$ and $(0, ~(i + F) \bmod Q)$ otherwise.

\section{Minimum Hop Count}\label{sec:min-hop-count}

In this section, we want to show how the geodetic positions of the satellites in the constellation can be leveraged to compute the minimum number of ISL-hops required to connect two satellites.
This section is based on previous work by Chen \etal in which a theoretical model to estimate the ISL hop count was presented~\cite{ChenGYFC21}.
We use these results as a foundation and augment them by our findings.
We do not consider how to select adequate access satellites for ground users but instead work with the assumption that the two satellites for which the hop count or route is to be calculated are given.

\subsection{Hop Count Model}\label{sec:mhc-formula}

In the following formulas, we often use $\lfloor\cdot\rceil$ to indicate that a number is rounded to the nearest integer according to the rounding function known as \emph{commercial rounding} that rounds half away from zero, \ie $\lfloor x\rceil = \sgn(x)\left\lfloor \vert x\vert + 0.5\right\rfloor$.
Note, however, that in our theoretical context, rounding is not strictly necessary as for all quotients the divisor is always a factor of the respective dividend.

Computing the minimum number of inter-plane hops~$H_h$ is fairly straightforward.
It is given by the horizontal distance that must be travelled to get from the initial orbital plane to the destination plane.
Therefore, it only depends on the difference between the longitudes of the respective ascending nodes:
\begin{align*}
    \Delta L_0 = (L_{0,2} - L_{0,1}) \bmod 2\pi \begingroup\color{tab-gray}\in \mathopen[0, 2\pi\mathclose[\endgroup
\end{align*}
$\Delta L_0$ is the longitudal angle that must be covered when going from the orbital plane of the source satellite to the plane of the destination in east direction.
Therefore, the RAAN difference in west direction is $2\pi - L_0$.
Since each hop from one plane to the next covers an angle of $\Delta \Omega$, the total number of inter-plane hops in east or west direction is given by:
\begin{align*}
    H_h^\leftarrow  & = \left\lfloor\frac{2\pi - \Delta L_0}{\Delta\Omega}\right\rceil \qquad &
    H_h^\rightarrow & = \left\lfloor\frac{\Delta L_0}{\Delta\Omega}\right\rceil
\end{align*}
Next, we need to look at the phase angle differences between the two satellites.
An intra-plane hop (to the successor satellite) adds $\Delta\Phi$ to the phase angle, while an inter-plane hop (towards the east) increases the phase angle by $\Delta f$.
Formally, when taking only hops to the successor and right neighbour, the following relationship holds:
\begin{align*}
    u_2 = u_1 + (H_h^\rightarrow \cdot \Delta f) + \underbrace{(H_v^\nearrow \cdot \Delta \Phi)}_{\Delta \overset{\rightarrow}{u}}
\end{align*}
Since we want to compute the number of intra-plane hops~$H_v$, we have to compute the fraction $\Delta \overset{\rightarrow}{u}$ of the phase angle difference that will be covered by intra-plane hops.
We again need to distinguish between two directions:
\begin{align*}
    \Delta \overset{\rightarrow}{u} & = (u_2 - u_1 - H_h^\rightarrow\cdot \Delta f) \bmod 2\pi \\
    \Delta \overset{\leftarrow}{u}  & = (u_2 - u_1 + H_h^\leftarrow\cdot \Delta f) \bmod 2\pi
\end{align*}
Finally, we can do similar calculations as for $H_h$ to compute the directional intra-plane hop counts:
\begin{align*}
    H_v^\nwarrow & = \left\lfloor\frac{\Delta \overset{\leftarrow}{u}}{\Delta\Phi}\right\rceil \quad        &
    H_v^\nearrow & = \left\lfloor\frac{\Delta \overset{\rightarrow}{u}}{\Delta\Phi}\right\rceil               \\
    H_v^\swarrow & = \left\lfloor\frac{2\pi - \Delta \overset{\leftarrow}{u}}{\Delta\Phi}\right\rceil \quad &
    H_v^\searrow & = \left\lfloor\frac{2\pi - \Delta \overset{\rightarrow}{u}}{\Delta\Phi}\right\rceil
\end{align*}
The minimum hop count is then just given by the minimum of the possible combinations:
\begin{equation*}
    \min\{H_h^\leftarrow+H_v^\nwarrow,~H_h^\leftarrow+H_v^\swarrow,~H_h^\rightarrow+H_v^\nearrow,~H_h^\rightarrow+H_v^\searrow\}
\end{equation*}
Note that this formula also contains indicators for the two directions in which to travel to achieve the minimum number of hops.

\subsection{Enhancement over Existing Formula}

\begin{figure}
    \centering%
    \input{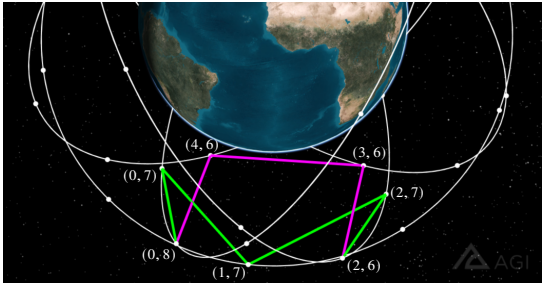}%
    \caption{A Walker Delta constellation $\ang{60}\colon 50/5/2$ pinpointing the difference where the original hop count formula produces suboptimal results.}
    \label{fig:incorrect-hopcount}
\end{figure}

As mentioned, the above derivations are inspired by results established by Chen \etal~\cite{ChenGYFC21}.
The latter are based on the assumption that \emph{``if the path on a given direction is too long (\eg $H_h > \sfrac{P}{2}$), the packets will go through the opposite direction.''}
The above results relinquish this assumption, because it may induce unnecessarily high hop counts.
As a concrete example, \autoref{fig:incorrect-hopcount} shows a Walker Delta constellation ($\ang{60}\colon 50/5/2$) that visualises this fine point.
The constellation contains $P = 5$ planes which means that according to the assumption, for all routes with more than two inter-plane hops, there should be an alternative route with $H_h \leq \sfrac{P}{2}$ and $H_v \leq \sfrac{Q}{2}$ that contains fewer or equally many hops.
Now consider the two routes from $(0, 8)$ to $(2, 6)$, for which the original work of Chen \etal~\cite{ChenGYFC21}, gives a minimum hop count of four.
The corresponding route is depicted in green.
It has one intra-plane hop, then two inter-plane hops and one more intra-plane hop.
The purple route has a hop count of only three (which is the value returned by our formula).
It consists of three inter-plane hops.
Notably, it is also shorter in total distance spanned.

\subsection{Hop Count Evaluation}\label{sec:mhc-comparison}

We validated the correctness of our formula by empirically comparing it to an algorithm that derives the hop count from the satellites' identifiers $(o, i)$.
Then, we evaluated the formula and analysed the differences on the Starlink constellation (see \autoref{sec:starlink}).
Across all possible combinations of satellite pairs, the hop count using Chen's formula was unnecessarily high in about $\SI{2.7}{\percent}$ of the pairs ($\SI{1.26}{\percent}$ one additional hop, $\SI{1.01}{\percent}$ three additional hops, $\SI{0.44}{\percent}$ five additional hops).

Furthermore, we compared the hop counts with those of the shortest-distance routes which in turn were computed using Dijkstra's algorithm.
Interestingly, the latter routes sometimes contain one more hop than the minimum ($\approx \SI{1}{\percent}$ of the pairs).
However, whenever this is the case, both hop count formulas agree on the count.

\section{Routing Algorithms}\label{sec:routing}

The second part of this paper considers different solutions for solving the routing problem.
First, we introduce Dijkstra's algorithm as the classical algorithm for solving shortest path problems.
Next, we argue why considering a metric based on hop count can be superior to just considering the length, \ie the total distance spanned by a route.
Finally, we suggest and evaluate new routing algorithms that integrate the minimum hop count.

\subsection{Dijkstra's Shortest Path}

\begin{algorithm}
    \caption{Dijkstra's Shortest Path}\label{alg:dijkstra}
    \small\begin{algorithmic}[1]
    \Procedure{Dijkstra}{$\mathit{Const.}, \mathit{src}, \mathit{dst}$}
    \ForAll{satellites $v \in \mathit{Constellation}$}
        \State $\mathit{dist}[v] \gets \infty$ \Comment{unknown distance to $v$}
        \State $\mathit{prev}[v] \gets \bot$ \Comment{predecessor of $v$}
        \State $\mathit{visited}[v] \gets \mathit{false}$
    \EndFor
    \State $\mathit{dist}[\mathit{src}] \gets 0$
    \State $Q \gets \Call{Heapify}{\{(v, \mathit{dist}[v]) \mid v \in \mathit{Const.}\}}$
    \While{$Q$ is not empty}
        \State $(u, d) \gets Q.\Call{Pop}{\null}$ \Comment{pop sat.\@ $u$ with min.\@ distance $d$}
        \State $\mathit{visited}[u] \gets \mathit{true}$
        \If{$u = \mathit{dst}$} \label{alg:dijkstra:line:shortcut}
            \State \textbf{break} \Comment{shortest path found}
        \EndIf \label{alg:dijkstra:line:shortcut2}
        \ForAll{neighbors $v$ of $u$}
            \If{$\neg \mathit{visited}[v]$}
                \State $\mathit{alt} \gets d + \Call{Euclidean}{u, v}$
                \If{$\mathit{alt} < \mathit{dist}[v]$}
                    \State $\mathit{dist}[v] \gets \mathit{alt}$ \Comment{found shorter alternative}
                    \State $\mathit{prev}[v] \gets u$
                    \State $Q.\Call{DecreaseKey}{v, \mathit{alt}}$ \Comment{update priority}
                \EndIf \begingroup\algrenewcommand\algorithmiccomment[1]{\hfill{\color{gray}#1}}\Comment{of satellite $v$}\endgroup
            \EndIf
        \EndFor
    \EndWhile
    \State \textbf{return} $\mathit{dist}[]$, $\mathit{prev}[]$
    \EndProcedure
\end{algorithmic}
\end{algorithm}

\begin{algorithm}
    \caption{Dijkstra's Path Reconstruction}\label{alg:dijkstra-reconstruct}
    \small\begin{algorithmic}[1]
    \Procedure{ReconstPath}{$\mathit{src}, \mathit{dst}, \mathit{prev}[]$}
    \State $S \gets [\mathit{dst}]$
    \State $u \gets \mathit{dst}$
    \While{$u \neq \mathit{src}$}
        \State $u \gets \mathit{prev}[u]$
        \State $S \gets u::S$ \Comment{insert $u$ at the front of $S$}
    \EndWhile
    \State \textbf{return} $S$
    \EndProcedure
\end{algorithmic}
\end{algorithm}

Dijkstra's algorithm~\cite{Dijkstra59} is a well-known algorithm and the de-facto standard for solving shortest path problems.
It is also commonly used in the domain of network routing and exists in various flavours.
In this paper, we are only interested in solving the single-pair shortest path problem.
We consider the Euclidean distance between satellites as a metric and use a min-heap to store the unvisited nodes.
The pseudocode is shown in \autoref{alg:dijkstra}.
Note that to compute the shortest paths from the source to all other nodes, it suffices to remove the lines \ref{alg:dijkstra:line:shortcut} to \ref{alg:dijkstra:line:shortcut2} which serve as a shortcut to terminate the algorithm once the destination node has been reached.

Once the algorithm finishes, the total distance of the shortest path from source to destination is $\mathit{dist}[\mathit{dst}]$.
However, we are usually also interested in the route itself rather than just its length.
Therefore, the code displayed in \autoref{alg:dijkstra} also remembers for each visited node its predecessor node on the shortest path to it.
\autoref{alg:dijkstra-reconstruct} shows how this $\mathit{prev}$ array can be used to reconstruct the shortest path.
The algorithm starts at the destination node and basically follows the pointers to the respective previous nodes until it arrives at the source node, keeping track of all nodes that were passed.

We use Dijkstra's algorithm in the following as a baseline for finding shortest routes between pairs of satellites and use it to compare the performance of other algorithms.

\subsection{Shortest Length vs.\texorpdfstring{\@}{} Minimum Hops}\label{sec:shortest-path-minimal-hops}

\begin{figure}
    \centering%
    \input{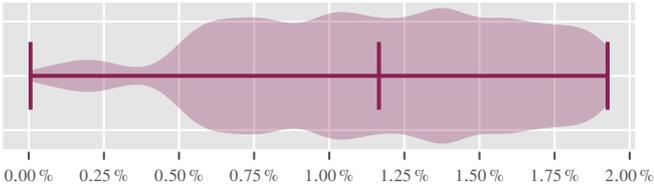}%
    \caption{Violin plot showing how much longer the shortest route with minimum number of hops is compared to the overall shortest route for Starlink.}\label{fig:shortest-path-additional-hop}
\end{figure}

In this section, we describe how the routing method can be optimised by exploiting the minimum hop count formula.
This improvement is based on the observation that the shortest route usually also has the lowest possible number of hops.
\autoref{sec:mhc-comparison} already showed that there are some exceptions to this where, in some rare cases, the shortest route can contain more hops than the minimum.
\autoref{fig:shortest-path-additional-hop} is a violin plot for the Starlink constellation which shows how much longer the shortest route with minimum number of hops is compared to the overall shortest path.
It considers only the ${\approx}\SI{1}{\percent}$ of satellite pairs where the shortest path has more hops than the minimum.
The largest difference is a ${\approx}\SI{2}{\percent}$ longer route (which amounts to around $\SI{1284}{\km}$ in absolute terms).
In all cases, there is only one additional hop.

However, we argue that the number of hops is actually the metric that should be minimised first.
Mega-constellations will need to carry large amounts of data, which is only possible with longer data packets, \eg jumbo Ethernet frames.
As the signals travel with the speed of light, it is clear that slightly shorter routes do not compensate for the overhead of an additional hop.
For example, the transmission time of a packet of size $\SI{65535}{bytes}$ (largest possible IPv4 payload) at an ISL rate of $\SI{1}{Gbps}$ is about $\SI{500}{\micro\second}$.
Even if faster ISL data rates are leveraged, there is internal packetisation and queuing delay which can easily add up to several milliseconds of on-board processing, even when using highly efficient ATM fabrics~\cite{DBLP:journals/comsur/CerovicPAHP18}.
A single hop is thus comparable with the propagation delay of $\SI{1000}{\km}$ distance at the speed of light ($\SI{299792}{\km\per\s}$) totalling $\SI{3.3}{\milli\second}$.

Knowing the minimum number of hops in both dimensions and their directions between two satellites allows the routing algorithm to restrict the exploration:
The search space can be reduced significantly, forming a spherical rectangle where the source and destination satellites are at opposing corners.
Note that in this rectangle, every possible route from the source to the destination has the same number of hops (similar to a Manhattan Street Network~\cite{DBLP:journals/tcom/Maxemchuk87}).

The fact that the search space is now a two-dimensional grid has the side-effect that it becomes a directed acyclic graph (DAG).
For DAGs, there is a more efficient algorithm than \textsc{Dijkstra}.
\textsc{DAGshort} is based on topological sorting and runs in $\mathcal{O}(V+E)$ time~\cite{DBLP:books/mg/CormenLR89}.
A topological order is a linear ordering of the graph's nodes where for every directed edge in the graph, the source is sorted before the destination in the ordering.
There exists a trivial ordering for the two-dimensional grid of ISL links:
Starting at the source node, all vertices in the grid can be enumerated line by line, \ie sorting them first by their vertical distance to the source and then by their horizontal distance.

For the sake of completeness, we want to mention yet another possible approach, namely a modification of \textsc{Dijkstra}, denoted \textsc{DijkstraHops}, that computes the shortest route amongst the ones with the minimum number of hops.
For this, it suffices to store tuples $(\mathit{hops}, \mathit{distance})$ on the heap and compare them using a lexicographic order.

\subsection{Probabilistic Routing}

We now want to quantify the difference between the best and worst possible route in such a spherical rectangle.
The question is whether it actually pays off to invest in compute-intensive routing algorithms or whether it suffices to (randomly) send the packet via any route with minimum number of hops.
For this, we analyse an algorithm \textsc{CoinFlipRoute} that, given a source and destination, first computes the minimum hop count and directions using the formula from \autoref{sec:mhc-formula}.
Then, it flips a coin at each satellite to randomly decide in which of the two directions the packet should continue.
The only difficulty is to keep track of the vertical and horizontal hops to restrict the route to the spherical rectangle.
When a packet has, for example, travelled all of its horizontal hops, the route is forced to be completed with the remaining number of vertical hops.

Notably, the approach of locally flipping a coin is not the same as enumerating all possible routes in the rectangle and selecting one of them uniformly at random.
This is because some satellites are restricted in their choice, and therefore the possible routes do not all have the same probability.

As we will see in the evaluation (\autoref{sec:evaluation}), the \textsc{CoinFlipRoute} approach turns out to perform quite well in practice and is very cheap to compute.
However, we present an algorithm in the next section that is still easy to compute but aims at calculating more reasonable routes.

\subsection{\texorpdfstring{\discoroute}{DisCoRoute}}

\begin{algorithm}
    \caption{\discoroute (Cases A2A \& D2D)}\label{alg:discoroute-a2a}
    \small\newcommand{\concat}{\mathbin{{+}\mspace{-8mu}{+}}}

\begin{algorithmic}[1]
    \Procedure{\discoroute{}A2A}{$\mathit{Const.}, \mathit{src}, \mathit{dst}$}
    \State $H_h, H_v \gets \Call{MinHopCount}{\mathit{Const.}, \mathit{src}, \mathit{dst}}$
    \State \Comment{w.\,l.\,o.\,g. $s_{0,0} = \mathit{src}$, $s_{H_h,H_v} = \mathit{dst}$, $\mathit{dst}$ is north east of $\mathit{src}$}
    \State $\mathit{route}_s \gets [\mathit{src}]$
    \State $\mathit{route}_t \gets [\mathit{dst}]$
    \State $i \gets 0$
    \State $j \gets H_h$
    \For{$H_h$ many times}
        \State $\mathit{reward}_s \gets \vert \varphi_{i, 0} + \varphi_{i+1, 0}\vert$
        \State $\mathit{reward}_t \gets \vert \varphi_{j, H_v} + \varphi_{j-1, H_v}\vert$
        \If{$\mathit{reward}_s < \mathit{reward}_t$}
            \State $\mathit{route}_t \gets s_{j-1, H_v} :: \mathit{route}_t$
            \State $j \gets j-1$
        \Else
            \State $\mathit{route}_s \gets \mathit{route}_s :: s_{i+1, 0}$
            \State $i \gets i+1$
        \EndIf
    \EndFor
    \State \textbf{assert} $i = j$ \Comment{$s_{i, 0}$ and $t_{j, H_v}$ are on same orbital plane}
    \If{$H_v = 0$} \Comment{is $\mathit{route}_t[0]$ also last element of $\mathit{route}_s$?}
        \State $\mathit{route}_t \gets \mathit{route}_t[1 {:}]$ \Comment{remove first element}
    \Else
        \State $\mathit{route}_s \gets \mathit{route}_s \concat [s_{i, 1}, \ldots, s_{i, H_v - 1}]$
    \EndIf
    \State \textbf{return} $\mathit{route}_s \concat \mathit{route}_t$
    \EndProcedure
\end{algorithmic}
\end{algorithm}%

\begin{algorithm}
    \caption{\discoroute (Cases A2D \& D2A)}\label{alg:discoroute-a2d}
    \small\newcommand{\concat}{\mathbin{{+}\mspace{-8mu}{+}}}

\begin{algorithmic}[1]
    \Procedure{\discoroute{}A2D}{$\mathit{Const.}, \mathit{src}, \mathit{dst}$}
    \State $H_h, H_v \gets \Call{MinHopCount}{\mathit{Const.}, \mathit{src}, \mathit{dst}}$
    \State \Comment{w.\,l.\,o.\,g. $s_{0,0} = \mathit{src}$, $s_{H_h,H_v} = \mathit{dst}$, $\mathit{dst}$ is north east of $\mathit{src}$}
    \State $\mathit{route}_s \gets [\mathit{src}]$
    \State $\mathit{route}_t \gets [\mathit{dst}]$
    \State $i \gets 0$
    \State $j \gets H_v$
    \For{$H_v$ many times}
        \State $\mathit{reward}_s \gets \vert \varphi_{0, i} + \varphi_{0, i+1}\vert$
        \State $\mathit{reward}_t \gets \vert \varphi_{H_h, j} + \varphi_{H_h, j-1}\vert$
        \If{$\mathit{reward}_s < \mathit{reward}_t$} \label{alg:discoroute-a2d:line:selection}
            \State $\mathit{route}_s \gets \mathit{route}_s :: s_{0, i+1}$
            \State $i \gets i+1$
        \Else
            \State $\mathit{route}_t \gets s_{H_h, j-1} :: \mathit{route}_t$
            \State $j \gets j-1$
        \EndIf
    \EndFor
    \State \textbf{assert} $i = j$ \Comment{$s_{0, i}$ and $t_{H_v, j}$ are reachable via horiz. hops}
    \If{$H_h = 0$} \Comment{is $\mathit{route}_t[0]$ also last element of $\mathit{route}_s$?}
        \State $\mathit{route}_t \gets \mathit{route}_t[1 {:}]$ \Comment{remove first element}
    \Else
        \State $\mathit{route}_s \gets \mathit{route}_s \concat [s_{1, i}, \ldots, s_{H_h - 1, i}]$
    \EndIf
    \State \textbf{return} $\mathit{route}_s \concat \mathit{route}_t$
    \EndProcedure
\end{algorithmic}
\end{algorithm}%

\begin{figure*}
    \centering%
    \input{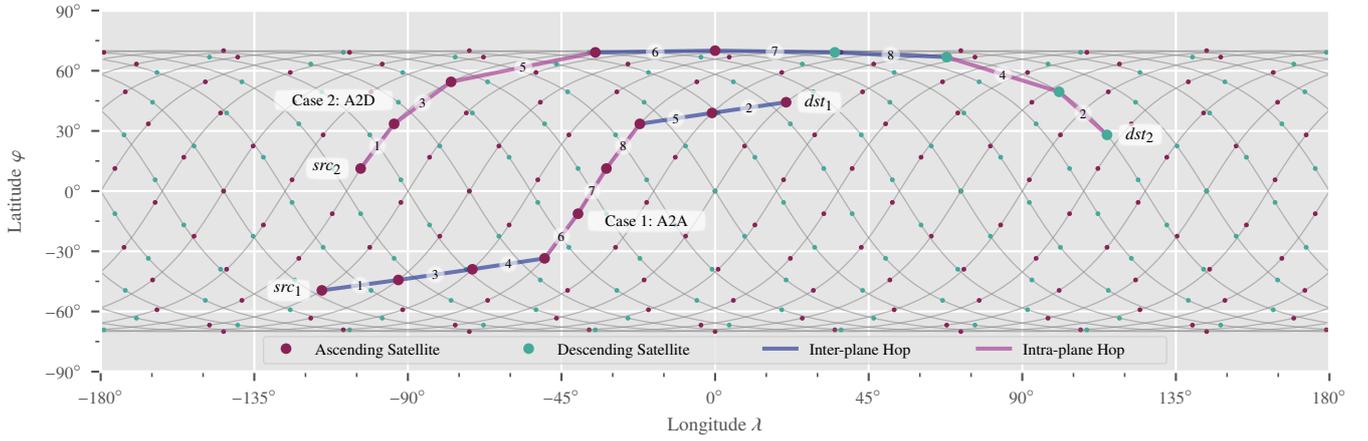}%
    \caption{Ground track of a Walker $\ang{70}\colon 300/20/5$ constellation showing an example route for each of the two cases of \discoroute.}
    \label{fig:discoroute-ground-plot}
\end{figure*}

\discoroute is a \textbf{dis}tributed routing algorithm for mega-\textbf{co}nstellations that exploits the insights found so far.
The key strength of this algorithm is that it is very cheap to compute compared to \textsc{Dijkstra}.
The algorithm produces near-optimal solutions that outperform the probabilistic approach.
In the following, we first introduce the idea and requirements of the algorithm, then describe how the algorithm works, and in \autoref{sec:evaluation}, we show that the loss of optimality is not significant in practice.
The algorithm is rooted in two insights:
\begin{enumerate}
    \item The length of an \emph{intra}-plane hop is always constant in the constellation, independent of the satellite's positions.
    \item The length of an \emph{inter}-plane hop should decrease the further it is away from the Equator.
\end{enumerate}
Therefore, the main idea is to cleverly distribute the inter-plane hops so that they happen as close as possible to the poles.

Similar to the probabilistic algorithm, we first need to compute the minimum hop count and the respective directions from source to destination.
Then, the algorithm distinguishes two cases depending on the flying direction of the two satellites.

\paragraph{Case 1: A2A / D2D}

Assume w.\,l.\,o.\,g.\@ that both satellites are ascending.
The idea is to distribute all inter-plane hops between the beginning and end of the route, having all intra-plane hops in the middle.
We chose the partition that maximises the overall distance of all inter-plane hops from the Equator.
The pseudocode is shown in \autoref{alg:discoroute-a2a}.
It starts by constructing the route simultaneously from the source and destination.
Then, it looks at the absolute value of the sum of latitudes of the inter-plane hop that starts at the source and of the one that ends at the destination.
The one with a larger value is added to the (respective) route.
This selection method is repeated for the number of horizontal hops many times.
Afterwards, the two route segments only need to be connected by the given number of intra-plane hops (potentially zero).

\paragraph{Case 2: A2D / D2A}

This case is roughly the inverse of the first case.
Routes from ascending to descending satellites must always go close to a pole since this is the only location where there is a link between an ascending and a descending satellite.
This means that the general rule should be to partition the intra-plane hops to the beginning and end of the route and have all inter-plane hops in the middle, as close to the pole as possible.
The pseudocode in \autoref{alg:discoroute-a2d} has a similar structure as the first case.
The only significant difference is the selection criterion in line~\ref{alg:discoroute-a2d:line:selection} that chooses an intra-plane hop on the side where the absolute value of the latitude sum of the involved satellites is smaller.

\autoref{fig:discoroute-ground-plot} shows an example for both cases.
The lower route corresponds to the first case.
It starts with an ascending satellite in the southern hemisphere and ends with an ascending satellite in the northern one, meaning that it must cross the Equator.
As one can see, the three inter-plane hops right at the beginning and the two at the end are distributed to maximise their overall distance from the Equator.
Note that if both satellites were in the northern hemisphere, the route would not start with inter-plane hops but would perform all of them at the end.
The route for the second case connects an ascending satellite with a descending one.
We can see that the selected intra-plane hops allow the inter-plane hops to have maximal distance from the Equator, \ie have the shortest length possible.

\section{Hop Distances}\label{sec:hop-distances}

This section covers the length calculations of inter- and intra-plane hops.
Using the following theorems, it becomes trivial to prove that the approximative \discoroute algorithm can solve the shortest path problem exactly for Walker Delta constellations with zero phase offset $\Delta f = 0$.

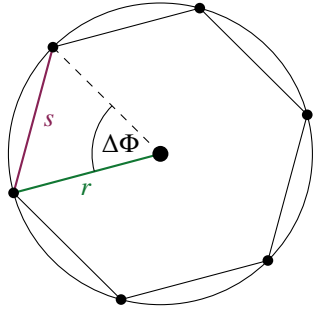
\begin{figure}
    \centering%
    \begin{tikzpicture}[]
    \pgfmathsetmacro{\r}{2}
    \draw (0, 0) coordinate (O) circle (\r);

    \pgfmathsetmacro{\DPHI}{360/6}
    \foreach \x in {0,...,5} {
        \coordinate (S\x) at ({\r * cos(\x * \DPHI + 15)}, {\r * sin(\x * \DPHI + 15)});
    }
    \draw (S0) -- (S1) -- (S2) (S3) -- (S4) -- (S5) -- (S0);

    \draw[tab-wine, thick] (S2) -- node[right] {$s$} (S3);
    \draw[tab-green, thick] (O) -- node[below] {$r$} (S3);
    \draw[dashed] (O) -- (S2);

    \pic["$\Delta\Phi$", draw, angle radius=9mm] {angle = S2--O--S3};

    \fill (O) circle (3pt);
    \foreach \x in {0,...,5} {
        \fill (S\x) circle (2pt);
    }
\end{tikzpicture}%
    \caption{Frontal view on an orbital plane with $Q = 6$ satellites, showing the circular orbit and 6-sided polygon.}
    \label{fig:circumradius}
\end{figure}

\begin{thm}[Length of Vertical Hops]\label{thm:distance-vertical-hop}
    The travel distance of an intra-plane hop is always the same, no matter where the hop is performed in the constellation.

    \begin{IEEEproof}
        Consider an orbital plane with $Q$ satellites.
        By construction, these satellites divide their circular orbit in circular arcs of equal length $\Delta\Phi = \sfrac{2\pi}{Q}$.
        See \autoref{fig:circumradius} for an example of an orbital plane with $Q = 6$ satellites.

        However, radio signals do not travel in arcs but in straight lines.
        The straight connection links between the satellites form a regular $Q$-sided polygon, where the circular orbit forms its circumscribed circle.

        The circumradius~$r$, which is the distance from the centre of the polygon to one of its vertices, is given by the satellite's altitude (measured from the centre of Earth).
        Therefore, the side length~$s$ of this polygon, which is the length of a direct link, is the base of an isosceles triangle with legs of length $r$.
        By the law of cosines, it holds that $s^2 = r^2 + r^2 - 2 r r \cos(\Delta\Phi)$.
        With the definition $\Delta\Phi = \frac{2\pi}{Q}$, it follows that $s = 2r \sin(\frac{\pi}{Q})$.
        This equation is only based on constants and is especially independent of the satellite position.
    \end{IEEEproof}
\end{thm}

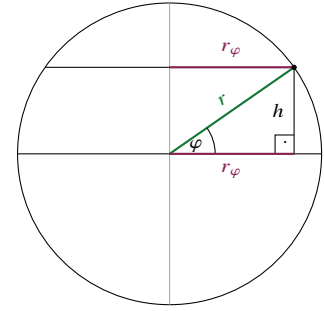
\begin{figure}
    \centering%
    \begin{tikzpicture}[scale=2, every node/.style={font=\scriptsize}]
    \draw (0, 0) coordinate (O) circle (1);
    \draw[tab-gray] (0, -1) -- (0, 1);
    \draw (-1, 0) -- (1, 0) coordinate (E);

    \draw ({-cos(35)}, {sin(35)}) -- ({cos(35)}, {sin(35)}) coordinate (S);
    \draw[tab-green, thick] (O) -- node[above, sloped] {$r$} (S);
    \draw (S) -- node[left] {$h$} (O-|S) coordinate (H);
    \pic["$\varphi$", draw, angle radius=6mm] {angle = E--O--S};
    \pic["$\cdot$", draw, angle eccentricity=.5, angle radius=2.5mm] {right angle = S--H--O};

    \draw[tab-wine, thick] (O) -- node[below] {$r_\varphi$} (H);
    \draw[tab-wine, thick] (S) -- node[above] {$r_\varphi$} (S-|O);

    \fill (S) circle (0.5pt);
\end{tikzpicture}%
    \caption{Frontal view on the Earth showing the Equatorial plane and a parallel disk at latitude~$\varphi$.}
    \label{fig:circumference-latitude}
\end{figure}

\begin{thm}[Length of Horizontal Hops]\label{thm:distance-horizontal-hop}
    In a constellation with zero phase offset $\Delta f = 0$, the travel distance of an inter-plane hop is shortest closest to the poles and longest around the Equator.

    \begin{IEEEproof}
        Consider a constellation with $\Delta f = 0$ and $P$ orbital planes.
        Since the phase offset is zero, horizontal neighbours always share the same latitude.
        This means that all $i$-th satellites of every orbital plane each form a disk parallel to the Equator.
        Independent of the latitude, the angular distance between two neighbours in a disk is always $\Delta\Omega = \sfrac{2\pi}{P}$.
        By the definition of cosine, the radius $r_\varphi$ of such a disk parallel to the Equator at latitude~$\varphi$ is given by $r_\varphi = r \cdot \cos\varphi$ (see \autoref{fig:circumference-latitude}).

        Similar to \autoref{thm:distance-vertical-hop}, each of these disks corresponds to the circumscribed circle of a regular $P$-sided polygon.
        Analogue to the previous proof, the side length, \ie the length of an inter-plane hop, is equal to $s_\varphi = 2r_\varphi \sin(\frac{\pi}{P}) = 2r \cos\varphi \sin(\frac{\pi}{P})$.
        This distance decreases towards the poles, \ie $\varphi \to \pm \ang{90}$ and is largest at the Equator.
    \end{IEEEproof}
\end{thm}

Our empirical evaluation suggests that \autoref{thm:distance-horizontal-hop} also applies to the more general case for non-zero phase offsets.
Unfortunately, we were unable to formally prove this and to provide a closed formula for the length of an inter-plane hop.
We nevertheless formulate our findings as two conjectures that describe when the inter-plane hop distance between two neighbouring satellites is maximal and minimal.

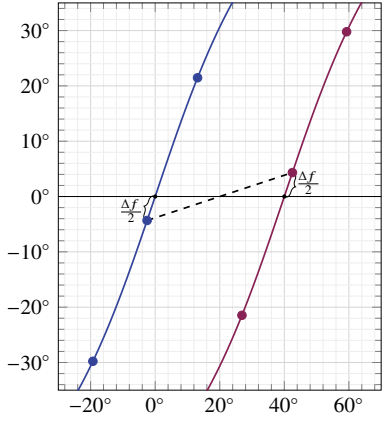
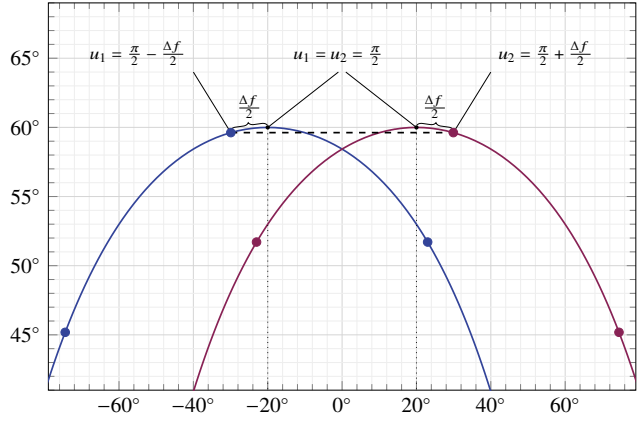
\begin{figure*}
    \centering%
    \begin{subfigure}{.38\textwidth}
        \centering%
        \begin{tikzpicture}[scale=0.8]
    \tikzmath{\inc = pi/3;}
    \begin{axis}[
            height=8cm, width=\linewidth, hide x axis=false, hide y axis=false,
            samples=500, domain=-pi/2:pi/2, trig format plots=rad,
            xmin=-30, xmax=70, ymin=-35, ymax=35,
            xticklabel={$\pgfmathprintnumber{\tick}^\circ$}, yticklabel={$\pgfmathprintnumber{\tick}^\circ$},
            grid=both, minor tick num=4, grid style={line width=.1pt, draw=tab-gray!20}, major grid style={line width=.2pt,draw=tab-gray!50}
        ]
        \addplot[tab-indigo, thick] plot ({mod((deg(atan(cos(\inc) * tan(\x)) + (pi/2 <= \x ? pi : 0) - (\x < -pi/2 ? pi : 0))) + 180, 360) - 180}, {deg(asin(sin(\inc) * sin(\x)))});
        \addplot[tab-wine, thick] plot ({mod((deg(atan(cos(\inc) * tan(\x)) + (pi/2 <= \x ? pi : 0) - (\x < -pi/2 ? pi : 0)) + 40) + 180, 360) - 180}, {deg(asin(sin(\inc) * sin(\x)))});

        \addplot[tab-indigo, only marks] coordinates {
            ({deg(atan(cos(\inc) * tan(rad(-35))))}, {deg(asin(sin(\inc) * sin(rad(-35))))})
            ({deg(atan(cos(\inc) * tan(rad(-5))))}, {deg(asin(sin(\inc) * sin(rad(-5))))})
            ({deg(atan(cos(\inc) * tan(rad(25))))}, {deg(asin(sin(\inc) * sin(rad(25))))})
        } coordinate [pos=0.5] (A1);
        
        \addplot[tab-wine, only marks] coordinates {
            ({deg(atan(cos(\inc) * tan(rad(-25)))) + 40}, {deg(asin(sin(\inc) * sin(rad(-25))))})
            ({deg(atan(cos(\inc) * tan(rad(5)))) + 40}, {deg(asin(sin(\inc) * sin(rad(5))))})
            ({deg(atan(cos(\inc) * tan(rad(35)))) + 40}, {deg(asin(sin(\inc) * sin(rad(35))))})
        } coordinate [pos=0.5] (B1);

        \draw[dashed, thick] (A1) -- (B1);
        \draw (-30, 0) -- (70, 0);
        \coordinate (AE) at (0, 0);
        \coordinate (BE) at (40, 0);

        \fill (AE) circle (1pt);
        \fill (BE) circle (1pt);

        \draw[decorate, decoration={brace, mirror, raise=1pt}] (AE) -- node[left, yshift=-1pt, font=\footnotesize] {$\frac{\Delta f}{2}$} (A1);
        \draw[decorate, decoration={brace, mirror, raise=1pt}] (BE) -- node[right, yshift=1pt, font=\footnotesize] {$\frac{\Delta f}{2}$} (B1);
    \end{axis}
\end{tikzpicture}%
        \caption{Maximal inter-plane hop distance.}
        \label{fig:horizontal-hop-maximal}
    \end{subfigure}%
    \begin{subfigure}{.62\textwidth}
        \centering%
        \begin{tikzpicture}[scale=0.8]
    \tikzmath{\inc = pi/3;}
    \begin{axis}[
            height=8cm, width=\linewidth, hide x axis=false, hide y axis=false,
            samples=500, domain=pi/4:3/4*pi, trig format plots=rad,
            xmin=-79, xmax=79, ymin=41, ymax=69,
            xticklabel={$\pgfmathprintnumber{\tick}^\circ$}, yticklabel={$\pgfmathprintnumber{\tick}^\circ$},
            grid=both, minor tick num=4, grid style={line width=.1pt, draw=tab-gray!20}, major grid style={line width=.2pt,draw=tab-gray!50}
        ]
        \addplot[tab-indigo, thick] plot ({mod((deg(atan(cos(\inc) * tan(\x)) + (pi/2 <= \x ? pi : 0) - (\x < -pi/2 ? pi : 0)) - 110) + 180, 360) - 180}, {deg(asin(sin(\inc) * sin(\x)))});
        \addplot[tab-wine, thick] plot ({mod((deg(atan(cos(\inc) * tan(\x)) + (pi/2 <= \x ? pi : 0) - (\x < -pi/2 ? pi : 0)) - 70) + 180, 360) - 180}, {deg(asin(sin(\inc) * sin(\x)))});

        \addplot[tab-indigo, only marks] coordinates {
            ({deg(atan(cos(\inc) * tan(rad(55)))) - 110}, {deg(asin(sin(\inc) * sin(rad(55))))})
            ({deg(atan(cos(\inc) * tan(rad(85)))) - 110}, {deg(asin(sin(\inc) * sin(rad(85))))})
            ({deg(atan(cos(\inc) * tan(rad(115))) + pi) - 110}, {deg(asin(sin(\inc) * sin(rad(115))))})
        } coordinate [pos=0.5] (A1);

        \addplot[tab-wine, only marks] coordinates {
            ({deg(atan(cos(\inc) * tan(rad(55 + 10)))) - 70}, {deg(asin(sin(\inc) * sin(rad(55 + 10))))})
            ({deg(atan(cos(\inc) * tan(rad(85 + 10))) + pi) - 70}, {deg(asin(sin(\inc) * sin(rad(85 + 10))))})
            ({deg(atan(cos(\inc) * tan(rad(115 + 10))) + pi) - 70}, {deg(asin(sin(\inc) * sin(rad(115 + 10))))})
        } coordinate [pos=0.5] (B1);

        \addplot[only marks, mark=.] coordinates {
            ({deg(atan(cos(\inc) * tan(rad(90)))) - 110}, {deg(asin(sin(\inc) * sin(rad(90))))})
            ({deg(atan(cos(\inc) * tan(rad(90)))) - 70}, {deg(asin(sin(\inc) * sin(rad(90))))})
        } coordinate [pos=0] (AX) coordinate [pos=1] (BX);

        \draw[dashed, thick] (A1) -- (B1);
        \draw[dotted] (-20, 40) -- (AX);
        \draw[dotted] (20, 40) -- (BX);

        \fill (AX) circle (1pt);
        \fill (BX) circle (1pt);

        \draw[decorate, decoration={brace, raise=1pt}] (A1) -- node[above, yshift=2pt, font=\footnotesize] {$\frac{\Delta f}{2}$} (AX);
        \draw[decorate, decoration={brace, mirror, raise=1pt}] (B1) -- node[above, yshift=2pt, font=\footnotesize] {$\frac{\Delta f}{2}$} (BX);
        
        \draw (A1) -- (-40, 64) node[anchor=south east, font=\footnotesize] {$u_1 = \frac{\pi}{2} - \frac{\Delta f}{2}$};
        \draw (B1) -- (40, 64) node[anchor=south west, font=\footnotesize] {$u_2 = \frac{\pi}{2} + \frac{\Delta f}{2}$};

        \draw (AX) -- (-1, 64) node[anchor=south, font=\footnotesize] {$u_1 = u_2 = \frac{\pi}{2}$};
        \draw (BX) -- (1, 64);
    \end{axis}
\end{tikzpicture}%
        \caption{Minimal inter-plane hop distance.}
        \label{fig:horizontal-hop-minimal}
    \end{subfigure}
    \caption{Satellite positions (in terms of phase angle~$u$) where the distance of an inter-plane hop is either (a)~maximal or (b)~minimal.}
\end{figure*}

\begin{conj}[Longest Horizontal Hop]
    An inter-plane hop between two neighbouring satellites has the longest distance when the phase angles of both satellites are equally distributed around the Equator.
    This means that the phase angles of the satellites are either $u_1 = -\sfrac{\Delta f}{2}$ and $u_2 = \sfrac{\Delta f}{2}$ or, analogously on the backside of the Earth, $u_1 = \pi - \sfrac{\Delta f}{2}$ and $u_2 = \pi + \sfrac{\Delta f}{2}$ (see \autoref{fig:horizontal-hop-maximal}).
\end{conj}

\begin{conj}[Shortest Horizontal Hop]
    A horizontal hop is smallest when the phase angles of both satellites are equally distributed around their points with maximal latitude.
    This means that the phase angles of the satellites are either $u_1 = \sfrac{\pi}{2} - \sfrac{\Delta f}{2}$ and $u_2 = \sfrac{\pi}{2} + \sfrac{\Delta f}{2}$ or, analogously on the backside of the Earth, $u_1 = -\sfrac{\pi}{2} - \sfrac{\Delta f}{2}$ and $u_2 = -\sfrac{\pi}{2} + \sfrac{\Delta f}{2}$ (see \autoref{fig:horizontal-hop-minimal}).
\end{conj}

It is noteworthy to mention that these two cases have a convenient property when considering the absolute value of the two satellites' latitudes $\vert u_1 + u_2\vert$ (as we do in \discoroute):
The longest distance of an inter-plane hop is achieved at the location where both satellites have the same latitude and their sum is maximal.
In contrast, the distance is minimal when the latitudes have the same magnitude but with different signs, \ie their sum is zero.

\section{Evaluation}\label{sec:evaluation}

In the last section of this paper, we provide the results of our empirical performance evaluation of the algorithms through simulations.
We have implemented our model and algorithms both in Python and Rust.
The measurements presented in this section are all produced using the Rust implementation as it is significantly faster than Python code.
However, this only affects the run time measurements since, apart from that, the two implementations calculate identical results.
All benchmarks were run on a Linux machine, equipped with an Intel Core{\footnotesize\texttrademark{}} \mbox{i7-6700} CPU running at $\SI{3.40}{\GHz}$ and $\SI{32}{\giga\byte}$ of main memory.

\subsection{Setup: Starlink Constellation}\label{sec:starlink}

Most of the following benchmarks are executed on the (first) orbital shell of the initial deployment phase of SpaceX's Starlink constellation.
More precisely, the Starlink constellation that we consider is formally described as a Walker Delta $\ang{53.0}\colon 1584 / 72 / 39$ at $\SI{550}{\km}$.
These parameters are taken from publicly available information~\cite{FCC:SAT-MOD-20190830-00087}, except for the phasing factor $F=39$ which is not explicitly mentioned and had to be estimated based on the available documents and by inspecting the publicly available data of satellites launched so far.

\subsection{Run Time}

\begin{figure}
    \centering%
    \input{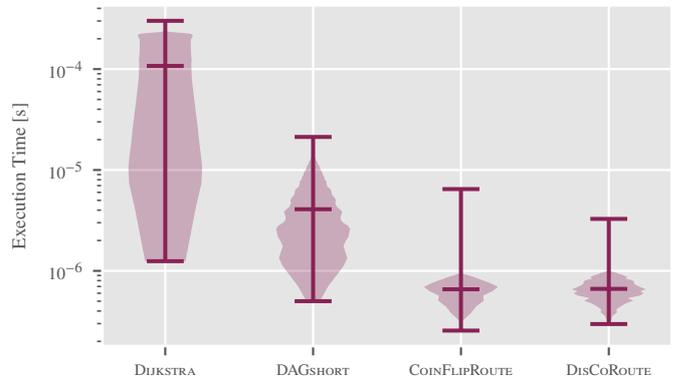}%
    \caption{Comparison of the execution times (in seconds) for each algorithm on all satellite pairs in the Starlink constellation.}
    \label{fig:bench-time-starlink}
\end{figure}

The violin plot in \autoref{fig:bench-time-starlink} shows the run time distribution of the different algorithms.
For each algorithm, we measure its execution time for all possible combinations of two satellites in the Starlink constellation.
Note that the plot uses a logarithmic scale and that we use the average of ten runs for each pair.
We observe that Dijkstra's algorithm is, as expected, the most expensive one.
This is mainly due to the fact that \textsc{Dijkstra} explores neighbour satellites in all four directions while all other algorithms first compute the minimum hop count and then restrict the search space to the induced spherical rectangle.
Comparing classical \textsc{Dijkstra} and \textsc{DAGshort}, we achieve a mean speedup of around $22\times$.
Considering \discoroute, we even end up with a mean speedup of $158\times$ compared to classical \textsc{Dijkstra} and still $7\times$ compared to \textsc{DAGshort}.
Interesting to note is that the \textsc{CoinFlipRoute} algorithm has a slightly lower mean than \discoroute but is slower in the worst case.

\subsection{Exactness}

\begin{figure}
    \centering%
    \input{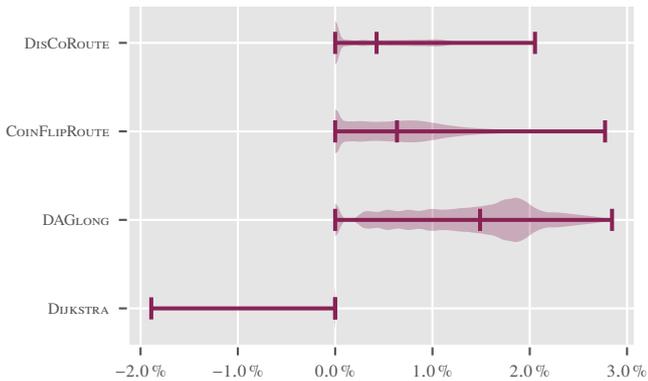}%
    \caption{Comparison of (relative) route lengths produced by each algorithm where the baseline is the shortest route with minimum hops, \ie \textsc{DAGshort}.}
    \label{fig:bench-starlink-length}
\end{figure}

Unlike \textsc{Dijkstra}, the hop-count-based algorithms are all approximates and do not always find the overall optimal solution.
This section is devoted to the analysis of the magnitude of suboptimality of these algorithms.
For this, we look again at all pairs in the Starlink constellation and compute the lengths of all computed routes.
We use the hop-count-based shortest path algorithm \textsc{DAGshort} as a baseline and compare the other algorithms relative to it.
This has the effect that classical \textsc{Dijkstra} is the only algorithm that can be better than the baseline.
For completeness, we also include an algorithm \textsc{DAGlong} that computes the length of the longest possible path inside the spherical rectangle induced by the hop counts.
The results are depicted in the violin plot in \autoref{fig:bench-starlink-length}.

As we have already seen in \autoref{sec:shortest-path-minimal-hops}, \textsc{Dijkstra} can outperform the baseline in only a tiny fraction of cases.
In the worst case, \textsc{Dijkstra} would be able to produce an up to $\SI{2}{\percent}$ shorter route.
However, the mean of \textsc{Dijkstra} is just at around $\SI{-0.02}{\percent}$.
Further, we see that \discoroute and \textsc{CoinFlipRoute} perform quite well in practice.

\subsection{Scalability}

\begin{figure}
    \centering%
    \input{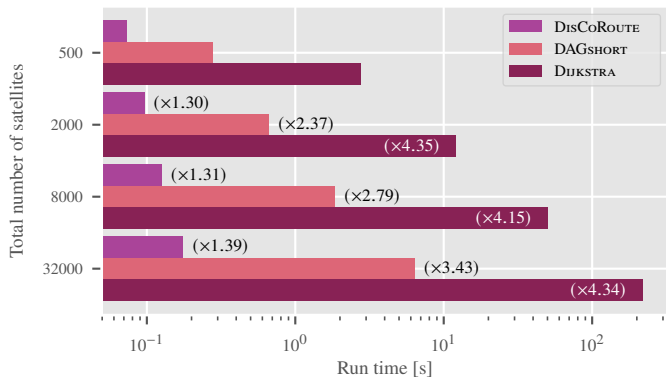}%
    \caption{Bar chart showing the scalability of the algorithms with increasing constellation size.}
    \label{fig:barchart-scalability}
\end{figure}

As a final performance indicator, we evaluate how well the different algorithms scale when the number of satellites in the constellation increases.
For this, we created four test-constellations $\ang{60}\colon 500/25/5$, $\ang{60}\colon 2000/50/10$, $\ang{60}\colon\num{8000}/100/20$, and $\ang{60}\colon\num{32000}/200/40$.
For each of them, we sampled $\num{100000}$ random satellite pairs and measured the execution time each algorithm takes.
The results are depicted on a logarithmic scale in the bar chart in \autoref{fig:barchart-scalability}.

Note that the number of satellites quadruples in each step.
The number in parentheses indicates the factor by which the run time increases compared to the previous case.
It is clear from the figure that our \discoroute algorithm provides better scalability on huge mega-constellations compared to \textsc{Dijkstra}.

\section{Conclusion}

This paper has started off with the question whether there exist simple routing schemes that optimise the number of hops as well as the travel distance between pairs of satellites in mega-constellations.
Apart from a straightforward probabilistic forwarding scheme, we have introduced the \discoroute algorithm and have presented profound empirical studies discussing the pros and cons in comparison to the optimal solution, thereby shedding light on the relative impact of travel time and hop count minimisation.
The results are overall very encouraging.
We have focussed on Walker Delta constellations (like Starlink) but are exploring extensions to other types of constellations.
Furthermore, we are currently embarking on extensions of this work, taking into account congestion and background traffic, as well as further evaluations (\eg route length and latency).

\section*{Acknowledgments}

This project has received funding from the European Union's Horizon 2020 research and innovation programme under the Marie Skłodowska-Curie grant agreement \href{https://cordis.europa.eu/project/id/101008233}{No 101008233} (\href{https://mission-project.eu}{MISSION}).

\printbibliography

\end{document}